\RequirePackage{rotating}
\documentclass[ijoc,nonblindrev]{informs3a} 

\OneAndAHalfSpacedXII 

\usepackage{natbib}
 \bibpunct[, ]{(}{)}{,}{a}{}{,}%
 \def\bibfont{\small}%
 \def\BIBand{and}%

\TheoremsNumberedThrough     
\def\TheoremsNumberedThrough{%
\theoremstyle{TH}%
\newtheorem{theorem}{Theorem}
\newtheorem{lemma}{Lemma}

\newtheorem{corollary}{Corollary}

\theoremstyle{EX}

}

\EquationsNumberedThrough    

\usepackage{subfigure,graphicx,caption,wrapfig}
\usepackage{booktabs}
\usepackage[linesnumbered, ruled]{algorithm2e}
\usepackage[usenames,dvipsnames,svgnames,table]{xcolor}
\usepackage{pgfplots}
\usepackage{ulem}

\usepackage{tikz}

\tikzstyle{vertex}=[circle,fill=black!15,minimum size=15pt,inner sep=0pt]
\tikzstyle{sourcesink}=[circle,fill=black!26,minimum size=20pt,inner sep=0pt]
\tikzstyle{midvertex}=[circle,fill=black!18,minimum size=20pt,inner sep=0pt]
\tikzstyle{selected vertex} = [vertex, fill=red!24]
\tikzstyle{edge} = [draw,thick,->]
\tikzstyle{weight} = [font=\small]
\tikzstyle{selected edge} = [draw,line width=4pt,->,red!50]
\tikzstyle{inserted edge} = [draw,thick, dashed,->]
\tikzstyle{ignored edge} = [draw,line width=3pt,-,black!20]

\newif\ifcomments
\commentstrue
\ifcomments
\newcommand{\kibitz}[2]{{\color{#1}{#2}}}
\else
\newcommand{\kibitz}[2]{}
\fi
\newcommand{\revout}[1]{\kibitz{red}{\sout{#1}}}
\renewcommand{\revout}[1]{}
\newcommand{\revin}[1]{\kibitz{black}{#1}}

\renewcommand{\theARTICLETOP}{}
\begin{document}
\normalem

\RUNAUTHOR{Benad\'{e} and Hooker}

\RUNTITLE{Optimization Bounds from the Branching Dual}

\TITLE{Optimization Bounds from the Branching Dual}

\ARTICLEAUTHORS{%
\AUTHOR{J.\ G.\ Benad\'{e}, J.\ N.\ Hooker}
\AFF{Carnegie Mellon University, \EMAIL{jbenade@andrew.cmu.edu}, \EMAIL{jh38@andrew.cmu.edu}}
} 

\ABSTRACT{%
We present a general method for obtaining strong bounds for discrete optimization problems that is based on a \mbox{concept} of branching duality.  It can be applied when no useful integer programming model is available, and we illustrate this with the minimum bandwidth problem.  The method strengthens a known bound for a given problem by formulating a dual problem whose feasible solutions are partial branching trees.  It solves the dual problem with a ``worst-bound'' local search heuristic that explores neighboring partial trees.  After proving some optimality properties of the heuristic, we show that it substantially improves known combinatorial bounds for the minimum bandwidth problem with a modest amount of computation.  It also obtains significantly tighter bounds than depth-first and breadth-first branching, demonstrating that the dual perspective can lead to better branching strategies when the object is to find valid bounds.
}%



\maketitle

%


\section{Introduction}

Establishing bounds on the optimal value of a problem is an essential tool for combinatorial optimization.  In a heuristic method, a good bound provides an indication of how close the solution is to optimality.  In an exact algorithm, a known bound can allow one to prove optimality of a feasible solution found early in the search.

We propose a general method for obtaining optimization bounds that is based on the concept of a {\em branching dual}.  It can, in particular, be applied to discrete optimization problems for which no useful integer programming models or cutting planes are available.  It begins with a known bound, perhaps a weak one, and builds a branching tree that strengthens the bound as much as desired.  

To obtain a good bound more quickly, we reconceive the branching process as local search in a dual space.  We regard partial branching trees as dual solutions of the optimization problem and obtain neighboring solutions by adding branches to the tree.  The value of a dual solution is defined to be the bound on the optimal value that is proved by the tree.  If the objective of the primal problem is to minimize, the dual problem seeks to maximize this bound.

This results in a different kind of branching scheme than ordinarily used in methods that seek an optimal solution.  Such methods typically attempt to solve both a primal and dual problem simultaneously.  They branch in such a way as to find good feasible solutions, while simultaneously seeking to prove a tight bound on the optimal value.  It is difficult to design a branching strategy that is effective at both tasks.  We propose instead to focus on the dual problem by constructing trees that are specifically designed to discover good bounds.  

The branching dual is clearly a strong dual, because a complete branching tree proves a bound equal to the optimal value.  In practice, however, we seek a suboptimal solution of the dual that yields a good bound after a reasonable amount of computation.  We do so by designing an effective local search procedure that takes advantage of problem structure.  This affords an alternative perspective that may yield a bound more quickly than standard branching procedures.


The approach is somewhat similar to a Lagrangian method in which bounds are obtained by partially solving the Lagrangian dual, perhaps by subgradient optimization.  Yet there are key differences.  Because there is no duality gap, the branching dual can deliver a bound as tight as desired if we invest sufficient computational resources.  Furthermore, there is no need for inequality constraints in the problem formulation \revin{(only inequality constraints can be dualized in a Lagrangian method)}, and no need to compute a subgradient or adjust the stepsize.

To solve the branching dual, we propose a {\em worst-bound} local search heuristic that examines  neighboring solutions obtained by branching at nodes with the worst relaxation value.  It is based on the principle that one should move to a neighboring solution that has some possibility of being better than the current solution.  

\revout{We will show that when the order in which one branches on variables is predetermined,}
\revin{We show that when the variable selected for branching at a node depends only on the node's level in the tree ({\em layered branching}),} the worst-bound heuristic is optimal in two senses.  It obtains any desired bound with a tree of minimum size, and it obtains the tightest possible bound that can be obtained from a tree of a given size.  In fact, these results hold more generally for \revin{\em fixed variable selection}, which means that the choice of branching variable at a node depends only on the choices along the path from the root to the node.   

When variable selection is not fixed, the heuristic examines neighboring solutions that result from various branching decisions.  The search can be designed to exploit the characteristics of the problem at hand, much as is done with local search methods in general.  We will see that a relatively simple local search procedure can substantially improve the bound.

The bound proved by a partial search tree is a function of the {\em relaxation values} computed at nodes of the tree.  The relaxation value at a node is a bound on the value of any solution obtained in a subtree rooted at that node.  If the problem has a tractable continuous relaxation, as in \revin{linear} integer programming, we can obtain a relaxation value simply by fixing the variables on which the search has branched so far and solving the continuous relaxation that results.  Relaxation values can often be obtained, however, without a continuous relaxation.  If there is a known combinatorial bound for a given problem, we need only determine how to alter the bound to reflect the fact that certain variables have been fixed.  This defines the relaxation values at nodes and allows a local search to improve the original bound, perhaps significantly.

We illustrate this strategy with the minimum bandwidth problem, for which no practical integer programming model is known.  Bounds for this problem have been studied at least since 1970, when 
Chv\'{a}tal introduced his famous density bound for the problem \citep{Chv70}.  Since the density bound is \mbox{NP-hard} to compute, polynomially computable bounds have been proposed, such as those of 
\citet{BluKonRavVem98} and 
\citet{CapSal05}.  We obtain relaxation values by adapting the Caprara\revin{--}Salazar-Gonz\'{a}lez bound to the case where some variables are fixed.  

We find in computational testing that the worst-bound heuristic delivers bounds that are not only better than the three bounds just mentioned, but that improve the Caprara\revin{--}Salazar-Gonz\'{a}lez bound significantly faster than depth-first and breadth-first branching trees that use the same relaxation values.  
\revout{We obtain these results both with a fixed branching order and without fixed variable selection.  In fact, when variable selection is not fixed in advance, a straightforward local search heuristic can significantly improve the bounds.}
\revin{We obtain these results both with a layered branching order,  where all the nodes on the same level branch on the same variable,  and without. In fact, when variable selection does not only depend on the level a node is on,  a straightforward local search heuristic can significantly improve the bounds.}
We conclude that the dual perspective proposed here can lead to better branching strategies when the object is to find valid bounds. 

The paper is organized a follows.  After a brief survey of related work, we define the branching dual and develop the idea of a relaxation function, which allows dual solutions to prove bounds on the optimal value.  We then describe the worst-bound heuristic and show that it is optimal for fixed variable selection.  The paper concludes with a computational study of the minimum bandwidth problem and remarks on future research.

\section{Related work}\label{sec:relatedwork}

A number of branching strategies have been proposed over the years, but almost always with the aim of solving a problem rather than obtaining a good dual bound quickly.  Depth-first search immediately probes to the bottom of the tree and may therefore discover feasible solutions early in the search.  It requires little space but tends to make slow progress toward improving the dual bound.  Breadth-first search explores all the nodes on one level before moving to the next.  It finds the best available bound down to the current depth but requires too much space for practical implementation.   

Primal/dual node selection strategies attempt to obtain some of the advantages of both depth-first and breadth-first search.  Iterative deepening \citep{Kor85} conducts complete depth-first searches to successively greater depths, each time re-starting the search.  It inherits the bound-proving capacity of breadth-first search while avoiding its exponential space requirement, but the amount of work still grows exponentially with the depth.  Limited discrepancy search \citep{HarGin95} conducts a depth-first search in a band of nodes of gradually increasing width.  Iteration 0 is a probe directly to the bottom of the tree.  Iteration $k$ is a depth-first search in which at most $k$ variables are set to values different from those in iteration 0.   This provides a bound at least as good as breadth-first search to level $k$, but the size of the search tree grows exponentially with $k$.  

Cost-based branching uses relaxation values at nodes as a guide to branching.  It is popular in mixed-integer solvers, where the relaxation values are obtained by solving (or estimating the solution value of) a continuous relaxation of the problem.  The two basic strategies are worst-first and best-first node selection.  Worst-first branching explores a node with the largest relaxation value first (if we are minimizing).  Strong branching \citep{AppBixChvCoo07,BixCooCoxLee95} might be viewed as similar to a worst-first strategy because it selects a branching variable that, when fixed, causes a large increase in the relaxation value.  Pseudocosts \citep{BenGauGirHenRibVib71,GauRib77} are often used instead of exact relaxation values to save computation time.  Worst-first branching is slow to improve the dual bound, because it leaves nodes with small relaxation values open longer.  This is of relatively little concern in branch-and-bound methods, because they use an upper bound and relaxation values at nodes (rather than the overall dual bound) to prune the search tree.  However, worst-first branching is a poor strategy for quickly obtaining a good dual bound.  Further discussion of these and related branching strategies can be found in \citet{AchKocMar05,Hooker12} and \citet{LinSav99}.

Best-first branching, by contrast, tends to improve the overall dual bound more quickly, because it explores nodes with the smallest relaxation value first.  It is nondeterministic because there may be multiple nodes with the same relaxation value.   It is shown in \citet{Ach07} that when variable selection is fixed, there exists a best-first node selection strategy that solves a given problem instance in a minimum number of nodes.  This, of course, leaves open the question of which best-first strategy achieves this result.  There is also the larger issue of which variable selection rule is best.

The worst-bound heuristic proposed here is based on the same idea as best-first branching but differs in that it simultaneously explores the children of all nodes with the smallest relaxation value.  We call it ``worst-bound'' rather than ``best-first'' to reflect this difference and our emphasis on the dual bound.  Because we are interested in bounding the optimal value rather than finding an optimal solution, we obtain somewhat stronger results than \citet{Ach07}.  Without assuming fixed variable selection, we show that for some selection of branching variables at nodes, the worst-bound heuristic proves any given valid bound with the minimum number of nodes.  This does not tell us which variable selection rule is best, but the heuristic can conduct a local search to find a promising variable to branch on at a given node.  Furthermore, we show that when the variable selection rule is fixed in advance, the worst-bound heuristic always proves a given bound with the minimum number of nodes, and it always proves the best bound that can be obtained by a tree with a given number of nodes.  

\revin{We therefore build on Achterberg's work in four ways: (a) we expand all relevant nodes with the minimum relaxation value, thus removing the non-determinism of best-first branching; (b) we prove the resulting algorithm is optimal for proving a dual bound; (c) we strengthen the algorithm with local search inspired by a concept of branching duality; and (d) we show empirically that the algorithm yields stronger dual bounds than depth-first and breadth-first branching.}

\revin{{\em Failure-directed search}, recently proposed by \citet{VilLabSha15} for scheduling problems, is similar to worst-bound branching in that it seeks to prove a bound (or infeasibility) rather than find a solution.  However, the mechanism is quite different, because it makes branching ``choices'' that are most likely to lead to infeasibility, based on the structure of the scheduling problem.  A ``choice'' is normally a higher-level decision, such as which currently unscheduled job to perform first.  {\em Orbital branching} \citep{OstLinRosSmr11} is designed for integer programming problems with a great deal of symmetry.  Groups of equivalent variables are used to partition the feasible region, so as to reduce the effects of symmetry.  It is unclear how these methods can be extended to a general branching method for optimization problems.}

The idea of branching duality was introduced for purposes of sensitivity analysis in \citet{Hoo96} and \citet{DawHoo00}.  It is further developed in \citet{Hooker12}, which suggests using a local search heuristic to solve the branching dual so as to obtain \revout{an}\revin{a} bound on the optimal value.  In the present paper, we carry out this suggestion by formulating a specific heuristic, proving its optimality properties, and applying it to the minimum bandwidth problem.

\section{The Branching Dual}\label{sec:elements}

The branching dual is most naturally defined for a problem with finite-domain variables.  We therefore consider an optimization problem of the form
\begin{equation}
    \min \;\{f(x)\;|\;x\in F,\; x\in D\}
    \label{eq:problem}
\end{equation}
where $x=(x_1,\ldots,x_n)$, $F$ is the feasible set, $D=D_{x_1}\times\cdots\times D_{x_n}$, and each $D_{x_j}$ is the finite domain of variable $x_j$.

A (partial or complete) {\em branching tree} for (\ref{eq:problem}) can be defined as follows.  Let $T$ be a rooted tree, and for any node $u$ of $T$, let $P[u]$ be the path from the root to node $u$.  We will say that $u$ is on {\em level $j$} of $T$ when $P[u]$ contains $j-1$ arcs.  A {\em terminal node} is any node on level $n+1$.  Then $T$ is a branching tree if 
\begin{description}
\item (a) every nonterminal node $u$ is {\em labeled} with a variable $x_{j(u)}$\revin{ designating the variable being branched on at $u$}, and the nodes in $P[u]$ have distinct labels;
\item (b) the arcs from any nonleaf node $u$ to its children are associated with distinct values in $x_{j(u)}$'s domain.  
\end{description}
The value associated with an arc leaving $u$ is viewed as an assignment to $x_{j(u)}$.  The arcs in $P[u]$ define a {\em partial assignment} $x[u]$ if $u$ is nonterminal and a {\em complete assignment} if $u$ is terminal.  

Each branching tree $T$ establishes a lower bound $\theta(T)$ on the optimal value of (\ref{eq:problem}), in a manner to be discussed in the next section.  We will regard the tree $T$ as a dual solution of (\ref{eq:problem}), and $\theta(T)$ as its value.  The {\em branching dual} of (\ref{eq:problem}) seeks a tree with maximum value:
\begin{equation}
\max \;\{\theta(T)\;|\;T\in\mathcal{T}\}
\label{eq:dual}
\end{equation}
where $\mathcal{T}$ is the set of branching trees for (\ref{eq:problem}).  The branching dual maximizes the bound that can be obtained from a branching tree.

\section{The Relaxation Function}

To relate the structure of a tree $T$ to the bound $\theta(T)$, we suppose that each node $u$ of $T$ has a {\em relaxation value} $c_u$.  This is a lower bound on the objective function value of any solution of (\ref{eq:problem}) consistent with the partial assignment $x[u]$.  We assume the following:
\begin{description}
\item (a) The relaxation value is nondecreasing with tree depth, so that $c_t\leq c_u$ when $t$ is a parent of $u$.
\item (b) The relaxation value is a sharp bound at any terminal node $u$, meaning that $c_u$ is exactly the value of the corresponding assignment $x[u]$.
\item (c) The relaxation value is a function solely of the partial assignment $x[u]$, so that we can write $c_u=c(x[u])$, where $c(\cdot)$ is the {\em relaxation function}.
\end{description}
Condition (c) is useful because it implies that the relaxation value of $u$ does not change when nodes are added to the tree.  This will allow us to prove various properties of the dual and algorithms for solving it.  

Let an {\em open node} $u$ of $T$ be a nonterminal node at which branching is still possible; that is, $u$ has fewer than $|D_{x_{j(u)}}|$ children.  A node that is not open is {\em closed}.  Thus we have the following.
\begin{lemma}  \label{le:bound}
A branching tree $T$ for (\ref{eq:problem}) establishes a bound $\theta(T)$ equal to the minimum of $c_u$ over all terminal and open nodes $u$ in $T$.  
\end{lemma}

The relaxation values can be obtained in any number of ways, so long as they satisfy (a)--(c).  They can be values of a linear programming relaxation, perhaps strengthened with cutting planes, or they can reflect combinatorial bounds, as in the discussion of the minimum bandwidth problem to follow.  They can also be strengthened by domain filtering and constraint propagation, as in constraint programming.  

We will assume that any feasibility checks are encoded in the relaxation value, so that $c_u=\infty$ whenever infeasibility is detected at node $u$.  We will say that $u$ is {\em infeasible} when $c_u=\infty$ and {\em feasible} when $c_u<\infty$.  An infeasible node is more accurately called a provably infeasible node, but for brevity we will refer to it simply as an infeasible node.

The branching dual is a strong dual because $\theta(T)$ is the optimal value of (\ref{eq:problem}) when $T$ is a complete branching tree.  $T$ is {\em complete} when every node of $T$ is closed or infeasible.

\begin{theorem}
If $T$ is a complete branching tree for (\ref{eq:problem}), then $\theta(T)$ is the optimal value of (\ref{eq:problem}).
\end{theorem}

{\em Proof.}  Suppose first that (\ref{eq:problem}) is feasible, and let $x^*$ be an optimal solution.  Let $P[u]$ be a longest path in $T$ for which $x[u]$ is consistent with $x^*$.  Suppose $u$ is nonterminal.  If $u$ is closed, some arc leaving $u$ assigns $x^*_{j(u)}$ to $x_{j(u)}$, which is impossible because $P[u]$ has maximal length.  Also $u$ cannot be infeasible, because $x^*$ is feasible.  Therefore, $u$ is terminal, which implies $c_u=\theta(T)=f(x^*)$.  If (\ref{eq:problem}) is infeasible and thus has value $\infty$,  any terminal node of $T$ must be infeasible.  Since any open node is infeasible, Lemma~\ref{le:bound} implies that $\theta(T)=\infty$. $\Box$
\medskip

\begin{corollary}
The branching dual is a strong dual.
\end{corollary}

\section{Solving the Dual}




The branching dual can be solved by a local search algorithm that moves from the current solution to a neighboring solution.  In general, a neighbor of $T$ could be any tree obtained by adding children to open nodes and/or removing leaf nodes.  We will suppose that the algorithm only adds nodes and does not remove them, because this prevents cycling and ensures that the number of iterations is bounded by the number of possible nodes.  In addition, the monotonicity of the relaxation function implies that the resulting dual values are nondecreasing.  Because there is no cycling, uphill search eventually finds an optimal solution.  \revin{This can still be regarded as local search in the sense that it searches a neighborhood of the current solution in each iteration.}

It remains to specify which nodes to add in each iteration.  Recall that the value of the current dual solution is governed by the worst (smallest) relaxation value of an open or terminal node $u$.  If $u$ is terminal, the heuristic terminates with the optimal bound $c_u$.   Otherwise, we propose adding nodes that can actually improve the current bound.  Expanding a node with \revin{a} relaxation value \revout{that is} better than the worst cannot improve the bound, because it leaves open nodes with relaxation values equal to the current bound.  However, expanding all nodes with the worst relaxation value can improve the bound.  We will refer to this as a {\em worst-bound heuristic}.

\revin{The heuristic is stated more precisely in Algorithm~1, in which $T$ is the current dual solution.}  An {\em eligible} node is an open node $u$ with relaxation value $c_u=\theta(T)$.  Note that every dual solution created by the heuristic is a {\em saturated} tree, meaning that all of its nonleaf nodes are closed.

\LinesNumberedHidden{
\begin{algorithm}
\caption{\revin{Worst-bound heuristic}}
\label{label}
\revin{
Let $T$ initially consist of the root node\;
\While{some open or terminal node in $T$ is feasible}{
  \eIf{some terminal node $u$ in $T$ has relaxation value $c_u=\theta(T)$}{
    stop with the optimal bound $\theta(T)$\;
  }{
    \For{each eligible node $u$ in $T$}{  
      select a label $x_{j(u)}$ for $u$ that does not occur in path $P[u]$\;
      add to $T$ all children of $u$ to create the next dual solution\;
    }
  }
}
problem (\ref{eq:problem}) is infeasible\;
}
\end{algorithm}
}

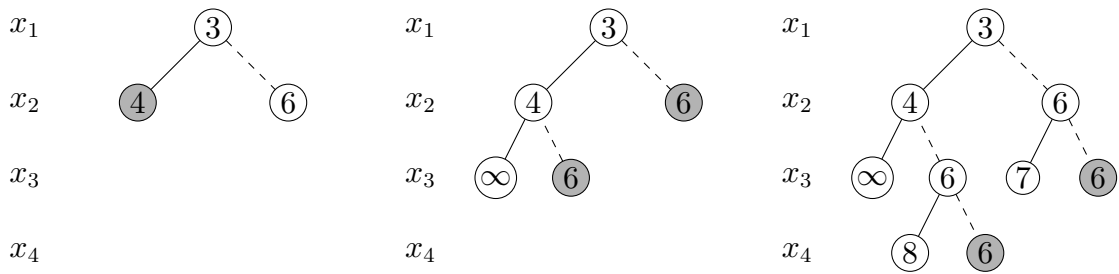
\begin{figure}[tb]
\centering
\begin{minipage}{.31\textwidth}
\centering
\begin{tikzpicture}
\node [draw,outer sep=0,inner sep=1.5,minimum size=10, shape=circle] (v1) at (-5.5,8.5) {3};
\node [draw,outer sep=0,inner sep=1.5,minimum size=10, shape=circle] (v3) at (-4.5,7.5) {6};
\node [draw,outer sep=0,inner sep=1.5,minimum size=10, shape=circle, fill=black!30] (v2) at (-6.5,7.5) {4};

\draw  (v1) edge (v2);
\draw [dashed] (v1) edge (v3);

\node at (-8,8.5) {$x_1$};
\node at (-8,7.5) {$x_2$};
\node at (-8,6.5) {$x_3$};
\node at (-8,5.5) {$x_4$};
\end{tikzpicture}
\end{minipage}
\begin{minipage}{.31\textwidth}
\centering
\begin{tikzpicture}
\node [draw,outer sep=0,inner sep=1.5,minimum size=10, shape=circle] (v1) at (-5.5,8.5) {3};
\node [draw,outer sep=0,inner sep=1.5,minimum size=10, shape=circle, fill=black!30] (v3) at (-4.5,7.5) {6};
\node [draw,outer sep=0,inner sep=1.5,minimum size=10, shape=circle] (v2) at (-6.5,7.5) {4};
\node [draw,outer sep=0,inner sep=1.,minimum size=10, shape=circle] (v21) at (-7,6.5) {$\infty$};
\node [draw,outer sep=0,inner sep=1.5,minimum size=10, shape=circle, fill=black!30] (v22) at (-6,6.5) {6};

\draw  (v1) edge (v2);
\draw [dashed] (v1) edge (v3);
\draw  (v2) edge (v21);
\draw [dashed] (v22) edge (v2);

\node at (-8,8.5) {$x_1$};
\node at (-8,7.5) {$x_2$};
\node at (-8,6.5) {$x_3$};
\node at (-8,5.5) {$x_4$};
\end{tikzpicture}

\end{minipage}
\begin{minipage}{.31\textwidth}
\centering
\begin{tikzpicture}

\node [draw,outer sep=0,inner sep=1.5,minimum size=10, shape=circle] (v1) at (-5.5,8.5) {3};

\node [draw,outer sep=0,inner sep=1.5,minimum size=10, shape=circle] (v3) at (-4.5,7.5) {6};
\node [draw,outer sep=0,inner sep=1.,minimum size=10, shape=circle] (v4) at (-5,6.5) {7};
\node [draw,outer sep=0,inner sep=1.5,minimum size=10, shape=circle, fill=black!30] (v55) at (-5.5,5.5) {6};
\node [draw,outer sep=0,inner sep=1.5,minimum size=10, shape=circle] (v6) at (-6.5,5.5) {8};
					
\node [draw,outer sep=0,inner sep=1.5,minimum size=10, shape=circle, fill=black!30] (v5) at (-4,6.5) {6};		

\node [draw,outer sep=0,inner sep=1.5,minimum size=10, shape=circle] (v2) at (-6.5,7.5) {4};
\node [draw,outer sep=0,inner sep=1.,minimum size=10, shape=circle] (v21) at (-7,6.5) {$\infty$};
\node [draw,outer sep=0,inner sep=1.5,minimum size=10, shape=circle] (v22) at (-6,6.5) {6};

\draw  (v1) edge (v2);
\draw [dashed] (v1) edge (v3);
\draw  (v2) edge (v21);
\draw [dashed] (v22) edge (v2);
\draw [dashed] (v22) edge (v55);
\draw [dashed]  (v3) edge (v5);
\draw  (v6) edge (v22);
\draw  (v3) edge (v4);
\node at (-8,8.5) {$x_1$};
\node at (-8,7.5) {$x_2$};
\node at (-8,6.5) {$x_3$};
\node at (-8,5.5) {$x_4$};
\end{tikzpicture}
\end{minipage}
\smallskip

\caption{Three iterations of the worst-bound heuristic for a layered variable ordering. Each node is inscribed with its relaxation value. All variables are binary:  a solid arc indicates assigning the value 1, a dashed arc 0. The shaded nodes will be examined in the next iteration.  }
\label{fig:tree}
\end{figure}

The heuristic must \revin{somehow} specify how to select a label for each eligible node $u$.  The labels are predetermined if variable selection is fixed, because in this case, the label at a node $u$ is a function of the labels on the other nodes along the path $P[u]$.  \revin{As an example, Fig.~\ref{fig:tree} shows how the worst-bound heuristic may proceed when variable selection is not only fixed, but branching is layered (i.e., each node on level $j$ receives label $x_j$).}

If variable section is not fixed, a local search is conducted to select labels for eligible nodes.  A greedy heuristic is the simplest approach, and we use it here.  For each eligible node $u$, examine a subset of the variables that are available to label $u$, and select one that will maximize the minimum relaxation value of $u$'s children.  The subset of variables considered depends on the characteristics of the problem at hand.  
\revin{Naturally, if the subset selected depends only on $P[u]$, the local search simply defines a fixed variable selection rule. However, if the subset is random or depends on factors other than $P[u]$, then variable selection is not fixed.  Even if the local search yields a fixed selection rule, a rule obtained at runtime may be better than one determined {\em a priori}.  We will find that this is in fact the case.  In addition, the worst-bound heuristic is optimal for a branching rule obtained at runtime when it is a fixed selection rule. }

The worst-bound heuristic is polynomial in the number of possible nodes, because the number of iterations is bounded by the number of nodes, and each iteration requires, at worst, examining each node of the current tree, and for each node, the children that result from selecting each possible label.

\section{Properties of the Worst-Bound Heuristic}

If variable selection is fixed, the worst-bound heuristic is optimal in two senses:  it finds the smallest branching tree that yields a given bound, and it finds the tightest possible bound that can be obtained from a tree of a given size.  We first establish a general result that holds even when there is no fixed variable selection.  We will say that branching tree $T'$ is a {\em branching subtree} of branching tree $T$ if $T'$ is a subtree of $T$ and the node labels in $T'$ are the same as in $T$.

\begin{theorem} \label{th:heuristic}
Given any branching tree $T$ for (\ref{eq:problem}) that establishes a bound $\lambda$, the worst-bound heuristic can be executed in such a way as to create a branching subtree of $T$ that establishes the same bound $\lambda$.
\end{theorem}

{\em Proof.}  We wish to show that the worst-bound heuristic can construct a branching subtree $T'$ of $T$ that establishes the bound $\lambda$.  We do so by first removing nodes from $T$ in a particular order until only the root node remains, and then constructing $T'$ by showing that the worst-first heuristic restores removed nodes in reverse order until $\lambda$ is proved.  We denote by $\lambda_1, \ldots, \lambda_k$ the distinct relaxation values of the nodes of $T$ that are less than or equal to $\lambda$, where  $\lambda_k=\lambda$ and $\lambda_1 < \cdots < \lambda_k$.  We next clean up $T$ by removing all leaf nodes whose parents have relaxation value of $\lambda_k$ or higher, and repeating until no such leaf nodes remain.  This yields a branching subtree $T_k$ of $T$ that still establishes bound $\lambda_k$.  Furthermore, $T_k$ is saturated, because if it contained an open nonleaf node $u$, then either $c_u<\lambda_k$ or $c_u\geq\lambda_k$.  In the former case, $T_k$ would not prove the bound $\lambda$, and in the latter case, $u$ would be a leaf node because its children would be removed.  Now remove from $T_k$ all leaf nodes whose parents have relaxation value $\lambda_{k-1}$, and repeat until no such nodes remain.  This yields a saturated tree $T_{k-1}$ that establishes the bound $\lambda_{k-1}$.  In similar fashion, remove nodes to obtain trees $T_{k-2}, \ldots, T_1$ ($T_1$ will consist of the root node only).  Since trees $T_1,\ldots,T_k$ are saturated, they can now be reconstructed by adding nodes according to the worst-bound heuristic, provided each node is given the label it has in $T_k$.  If we let $T'=T_k$, $T'$ proves bound $\lambda$ and is a branching subtree of $T$, and the theorem follows.  $\Box$
\medskip

Theorem~\ref{th:heuristic} provides no guidance on how the heuristic should label nodes to obtain a desired bound.   However, if variable selection is fixed, the labels are determined by previous branches, as noted in the previous section.  In this case, we have the following.

\begin{corollary}\label{cor:nodelimit}
Suppose variable selection is fixed, and the worst-bound heuristic is terminated at a point where the current tree contains $N$ nodes.  This tree establishes the tightest bound that can be obtained from a tree of $N$ nodes.
\end{corollary}

{\em Proof.}  Suppose $T$ is the tree obtained from the worst-bound heuristic, and $T'$ is a tree of size $N$ that establishes a bound $\theta(T')>\theta(T)$.  By Theorem~\ref{th:heuristic}, the worst-bound heuristic yields a branching subtree $T''$ of $T'$ that establishes the bound $\theta(T')$, so that $\theta(T'')\geq\theta(T')$.  But since variable selection is fixed and the size of $T''$ is at most $N$, $T''$ is a branching subtree of $T$.  So we have $\theta(T)\geq \theta(T'')\geq \theta(T')>\theta(T)$, a contradiction.  $\Box$
\medskip

\begin{corollary}  Suppose variable selection is fixed, and the worst-bound heuristic is terminated as soon as it proves a bound of $\lambda$.  The resulting tree is the smallest tree that establishes the bound $\lambda$.
\end{corollary}

{\em Proof.}  Suppose $T$ is the tree obtained from the heuristic, and $T'$ is a smaller tree with $\theta(T')=\lambda$.  By Theorem~\ref{th:heuristic}, the worst-bound heuristic yields a branching subtree $T''$ of $T'$ that establishes the bound $\lambda$, for a suitable choice of node labels.  But since variable selection is fixed, $T$ must be a branching subtree of $T''$, because it is the first tree obtained by the worst-bound heuristic that proves $\lambda$.  Thus if we let size$(T)$ denote the size of $T$, we have size$(T)>\mathrm{size}(T')\geq\mathrm{size}(T'')\geq\mathrm{size}(T)$, a contradiction.  $\Box$
\medskip




\section{The Minimum Bandwidth Problem}\label{sec:computation}

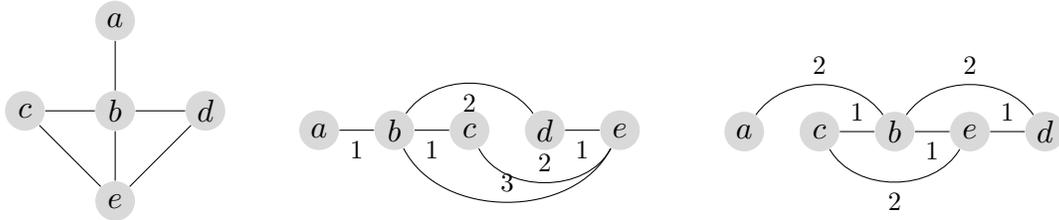
\begin{figure}[ht]\center  
\begin{tikzpicture}[scale=1.2, auto,swap]
    \foreach \y/\x/\name in {1/2/a,0/1/c,0/2/b,0/3/d,-1/2/e}
	    \node[vertex] (\name) at (\x,\y) [] {$\name$};
   
     \foreach \source/\dest/\weight in {a/b,b/c,b/d,b/e,d/e,c/e}
        \path[draw]   (\source)   to   node[weight] {} (\dest);
\end{tikzpicture}$\qquad$
\begin{tikzpicture}[scale=1, auto,swap]
    \foreach \y/\x/\name in {0/0/a,0/1/b,0/2/c,0/3/d,0/4/e}
	    \node[vertex] (\name) at (\x,\y) [] {$\name$};   
     \foreach \source/\dest/\weight in {a/b/1,d/e/1,b/c/1}
        \path[draw]   (\source)   to   node[weight] {$\weight$} (\dest);
     \foreach \source/\dest/\weight in {b/d/2}
        \path[draw]   (\source)   to[bend left =55]   node[weight] {$\weight$} (\dest);
        \foreach \source/\dest/\weight in {c/e/2}
        \path[draw]   (\source)   to[bend right=63]   node[weight, swap] {$\weight$} (\dest);
        \foreach \source/\dest/\weight in {b/e/3}
        \path[draw]   (\source)   to[bend right=63]   node[weight, swap] {$\weight$} (\dest);
\end{tikzpicture} $\qquad$
\begin{tikzpicture}[scale=1, auto,swap]
    \foreach \y/\x/\name in {0/0/a,0/1/c,0/2/b,0/3/e,0/4/d}
	    \node[vertex] (\name) at (\x,\y) [] {$\name$};
   
     \foreach \source/\dest/\weight in {b/c/1, b/e/1,d/e/1}
        \path[draw]   (\source)   to   node[weight] {$\weight$} (\dest);
     \foreach \source/\dest/\weight in {a/b/2,b/d/2}
        \path[draw]   (\source)   to[bend left =55]   node[weight, swap] {$\weight$} (\dest);
        \foreach \source/\dest/\weight in {c/e/2}
        \path[draw]   (\source)   to[bend right=60]   node[weight] {$\weight$} (\dest);
\end{tikzpicture}
\caption{A graph and  two linear arrangements of its vertices with   edge lengths indicated.}\label{fig:minband}
\end{figure}

The minimum bandwidth problem asks for a linear arrangement of the vertices of a graph that minimizes the length of the longest edge, where the length of an edge is measured by the distance it spans in the arrangement.  That is, given a graph $G=(V,E)$, the problem is to find 
\begin{equation}
\phi(G) = \min_{\tau} \max_{(i,j) \in E} |\tau_i - \tau_j|
\label{eq:bandwidth}
\end{equation}
where $\tau$ is any permutation of $1, \ldots, |V|$, and $\tau_i$ is the position of vertex $i$ in the arrangement.
A graph with five vertices may be seen in Fig.~\ref{fig:minband}, together with two of its linear arrangements. The first linear arrangement has value 3 while the second has value 2 and is optimal.

Several graph theoretic lower bounds have been derived for this problem.  
Perhaps the best known is the \emph{density bound} of 
\citet{Chv70}.  Let $d(s,t)$ denote the distance between vertices $s,t \in V$, defined as the length of a shortest path connecting $s$ and $t$, where the length of a path is the number of edges in it.   Given vertex sets $S,T\subseteq V$, let the distance from $S$ to $T$ be $d(S,T) = \max\{d(s,t): s\in S, t\in T\}$, and let $d(S) = d(S,S)$ be the \emph{diameter} of $S$.  The density bound is defined to be
\begin{equation} \label{eq:density}
\beta(G) = \max_{S\subseteq V}\left\lceil\frac{|S|-1}{d(S)}\right\rceil = \max_{S\subseteq V}\min_{v\in S} \left\lceil \frac{|S|-1}{d(v,S)}\right \rceil.
\end{equation}
It is clear from the reformulation in (\ref{eq:density}) that calculating $\beta(G)$ is equivalent to finding the largest clique in $G$ and is therefore NP-hard. 

\citet{BluKonRavVem98} propose a $1/2$-approximation of the density bound,
\begin{equation}\label{eq:blumbound}
\alpha(G) = \max_{v\in V}\max_{\substack{S\subseteq V\\v\in S}} \left\lceil\frac{|S|-1}{2d(v,S)}\right\rceil = \max_{v\in V}\max_{k=1}^{d(v,V)} \left\lceil \frac{|N_k(v)|-1}{2k}\right \rceil,
\end{equation}
where $N_k(v) = \{u \in V: d(u,v) \leq k \}$ is the $k$-neighborhood of $v$. The reformulation shows that $\alpha(G)$ is computable in time $O(nm)$ by viewing every vertex as the root of a layered graph.

\citet{CapSal05} similarly propose a bound  computable through layered graphs,
\begin{equation}
\gamma(G) = \min_{v\in V}\max_{\substack{S\subseteq V\\v\in S}} \left\lceil\frac{|S|-1}{d(v,S)}\right\rceil = \min_{v\in V}\max_{k=1}^{d(v,V)} \left\lceil \frac{|N_k(v)|-1}{k}\right \rceil.
\label{eq:csbound}
\end{equation} 
It places $v$ in the first position of an arrangement and greedily places all the vertices in $N_k(v)$ directly after it in the arrangement.  This bound can also be computed in polynomial time.  As Caprara and Salazar-Gonz\'{a}lez point out, this bound has the advantage that it can be naturally be adapted to the case when some vertices have fixed positions, as in a branching tree.  We will exploit this advantage in the next section.



\section{Branching Dual for Minimum Bandwidth}

To formulate the branching dual of the minimum bandwidth problem, it is convenient to state the problem using different variables than in (\ref{eq:bandwidth}).  We let $x_i$ be the vertex that is placed in position $i$ of the arrangement.  Then the problem is
\begin{equation}
\min_{x} \hspace{-1ex} \max_{\substack{i,j\\(x_i,x_j)\in E}}\hspace{-1.5ex} |i-j|
\label{eq:bandwidth2}
\end{equation}
where $x=(x_1,\ldots, x_n)$ is a permutation of $1, \ldots, n$ and $n=|V|$.

We construct branching trees by branching on the variables $x_i$.  We use a branching order that alternates between assigning vertices to the left and right ends of the arrangement, because this will be convenient for the relaxation function described below.  Thus at each node $t$ in level $i$, the partial assignment $x(t)$ fixes variables $x_1,x_n,x_2,x_{n-1},x_3,x_{n-2}, \ldots, x_i$.  Let $L$ be the set of vertices that are assigned to the left end \revout{ (i.e., assigned by fixed variables $x_1,x_2,\ldots$), and $R$ the set of vertices assigned to the right end (i.e., assigned by fixed variables $x_n,x_{n-1},\ldots$). } \revin{of the arrangement by fixing vertices $u_1, \ldots, u_{|L|}$ to positions $1, \ldots, |L|$. $R$ is similarly the vertices fixed to the the right end of the arrangement with vertices $v_1, \ldots, v_{|R|}$ in positions $n, \ldots, n - |R| + 1.$} 
\revin{Denote with $F = V \setminus (L\cup R)$ the set of unplaced vertices.}

The relaxation value is defined as follows.  Since the vertices in $L\cup R$ have already been assigned positions, each arc leaving $t$ that assigns one of these vertices to $x_i$ leads to an infeasible child node $u$ with relaxation value $c_u=\infty$.  The remaining arcs lead to feasible child nodes.  To define the relaxation value $c_u$ at such a node, we modify the bound (\ref{eq:csbound}) as in \citet{CapSal05} to reflect the fact that vertices in $L$ and $R$ have been assigned positions.
The resulting value $c_u$ is given by the optimal value of the integer programming problem (29) stated in their article \revin{which we repeat here for completeness}.
\revin{
Let $\Pi$ be the set of all permutations of $\{1, \ldots, n\}$, and $\Pi_p$ the set of all permutations of subsets of $\{1, \ldots, n\}$ of cardinality $p$. The bi-level integer linear programming relaxation used as relaxation value is 
\begin{align*}
    & \min  \phi \\
    & f_v \leq \tau_v \leq \ell_v, \;\; v\in F,\\
    & \tau_{u_h} = h, \;\;  h = 1, \ldots, |L|, \\
    & \tau_{v_i} =  n - i + 1 , \;\; i = 1, \ldots, |R|, \\
    & \tau \in  \Pi \\ 
    & \ell_{u_h} = h, \;\;  h = 1, \ldots, |L|, \\
    & \phi \geq i - h, \;\; (u_i,u_h) \in E, \\
    & f_{v_i}  = n - i + 1 , \;\; i = 1, \ldots, |R|, \\
    & \phi \geq i-h, \;\; (v_i, v_h) \in E,  
\end{align*}
where 
\begin{align*}
& \ell_v = \max_{\pi^v\in \Pi_{|N_1^L(v)\cup\{v\}|}} \Big\{ \pi_v^v \; \Big| \;
\phi\geq \pi_v^v - \pi_u^v, \; \pi_u^v \leq \ell_u, \; \forall u\in N_1^L(v) \Big\}, \; v\in F \\
& f_v = \max_{\rho^v\in \Pi_{|N_1^R(v)\cup\{v\}|}} \Big\{ \rho_v^v \; \Big| \;
\phi\geq \rho_v^v - \rho_u^v, \; \rho_u^v \leq f_u, \; \forall u\in N_1^R(v) \Big\}, \; v\in F
\end{align*}
Here $N_1^L(v)$ is the set of nodes adjacent to $v$ for which the shortest distance to any node in $L$ is exactly one unit shorter than the shortest distance from $v$ to a node in $L$. $N_1^R$ is defined analogously.

The inner-level ILP's compute the accurate values for $f_v$ and  $\ell_v$, the first and last positions available to $v\in F$ without violating the current value of $\phi.$ The outer level ILP optimizes the bandwidth and location of the free vertices subject to these constraints on their positions while ensuring that the fixed vertices are placed in the correct locations and ensuring that the bandwidth is at least the length of the longest edge between two vertices in $L$ or $R$. Edges incident to a free vertex are implicitly considered when computing $\ell_v$ and $f_v$, but edges between a vertex in  $L$ and one in  $R$ do not affect the bandwidth, making this a relaxation. 

Propositions~12 and~15 in \cite{CapSal05} present a simple algorithm for solving this problem which performs binary search on the value of $\phi$ guided by the feasibility of the bounds imposed on the unfixed vertices by $f_v$ and $\ell_v$.  The complexity of the algorithm is $O(m\log n + n\log^2 n)$.
}
\revout{
  In addition, Propositions~12 and~15   present a simple algorithm for solving the problem.  It computes the first and last feasible position of every unfixed vertex $w$ for a given $\phi$, denoted $f_w$ and $\ell_w$ respectively, by taking the relationship between the fixed and unfixed vertices into account.  Depending on whether or not it is possible to find an arrangement satisfying these constraints, the algorithm decreases or increases $\phi$ as in a binary search procedure.  The complexity of the algorithm is $O(m\log n + n\log^2 n)$.
  }

Consider, as a simple example, the graph in Figure \ref{fig:minband}. Suppose that $\phi \leq 2$ and vertex $c$ has been fixed \revin{to} the first position of the arrangement, so $\tau_c = 1$. This allows us to bound the domains of its neighbours, $\ell_b = 3$ and $\ell_e = 3$ and similarly $\ell_d = 5$, which of course has been known all along. Observe that one of vertices $b$ or $e$ must be placed into position 2, so although $\ell_c = \ell_e = 3$, we improve on the bounds by inferring $\ell_d = 4$ followed by  $\tau_a =5$.


We can also include variable selection in the worst-bound heuristic, using the greedy algorithm described earlier, rather than relying on a fixed\revin{, alternating} branching order. We consider only two candidates for the next variable on which to branch.  At each eligible node $u$, we let the branching variable be the next variable on the left or the next variable on the right, rather than strictly alternating as above.  Thus if the currently fixed variables at $u$ are $x_1, \ldots, x_i$ and $x_k, \ldots, x_n$, we choose between $x_{i+1}$ and $x_{k-1}$ as the next branching variable.  The greedy choice is the variable that maximizes the smallest relaxation value among $u$'s children.

\section{Computational Results}

We compare bounds obtained by the worst-bound heuristic (WBH) to known graph-theoretic bounds for the minimum bandwidth, as well as to bounds obtained from depth-first search (DFS) and breadth-first search (BFS).   We find that the WBH can strengthen bounds more rapidly than DFS and BFS when a fixed\revin{, layered} branching order is used.  Furthermore, these bounds can be improved significantly when variable selection is included in the heuristic, rather than using a fixed \revin{layered} branching order.

We use two types of randomly generated test instances\revin{ and a set of benchmark instances from the literature}. The first \revout{type}\revin{set of randomized instances}, denoted \texttt{Random}, consists of 90 random graph instances with 30 vertices.  We generated 10 instances for each density $ d\in \{0.1, \ldots, 0.9\}$ \revout{The instances are generated} so that every edge independently has probability $d$ of occurring.  

The second type of instances, denoted \texttt{Turner}, are generated according to the random model of \citet{Tur86}, which was also used for experiments in \citet{CapSal05}. This model controls the bandwidth to be at most $\phi$ while keeping the density fixed at $d\in \{0.3,0.5\}$.  We generated instances with 30 vertices for $\phi \in  \{3, 6, \ldots, 27\}$, resulting in 180 instances, \revout{and} instances with 100 vertices for $\phi \in \{10,20,30,40,50\}$, \revout{resulting in another 100 instances.} \revin{and instances with 250 and $1\,000$ vertices for $\phi \in\{20,40,60,80\}$ and $\phi \in \{50,100,150,200\}$, respectively, for a total of 440 instances.} 

\revin{Finally, we test on the set of \emph{Matrix Market}\footnote{Available at \url{http://math.nist.gov/MatrixMarket/}.} benchmark instances. As in \citet{CapSal05}, we restrict ourselves to instances with up to  250 vertices, leaving 39 instances with between 24 and 245 vertices (including 26 with at least 100 vertices). }

\begin{figure}[tb]
\caption{\revin{A performance profile showing the fraction of  \texttt{Random} instances with 30 vertices for which a target bound is proved. The blue baseline is the fraction of instances where the unstrengthened graph theoretic bounds achieved the target bound.  The partial tree is limited to   $k=100$ or $k=10,000$ internal nodes.} }\label{fig:rand30}
\includegraphics[scale=.42]{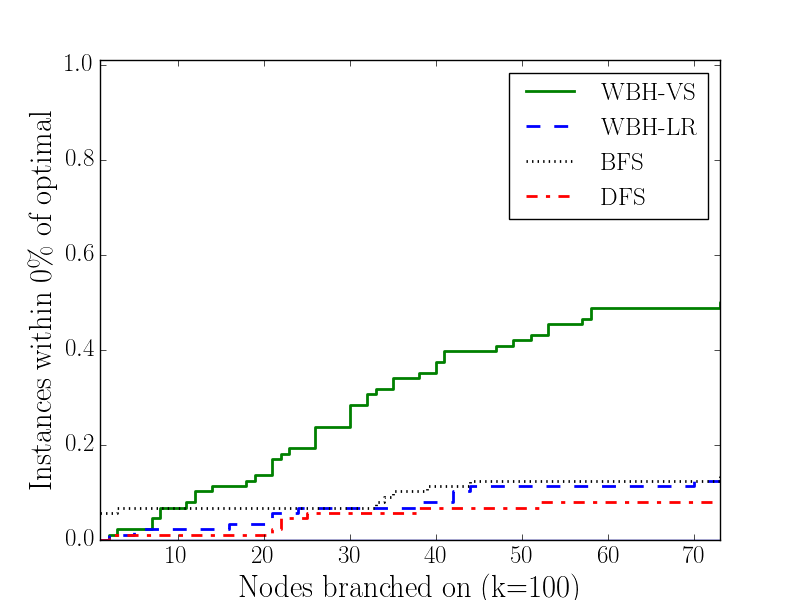}
\includegraphics[scale=.42]{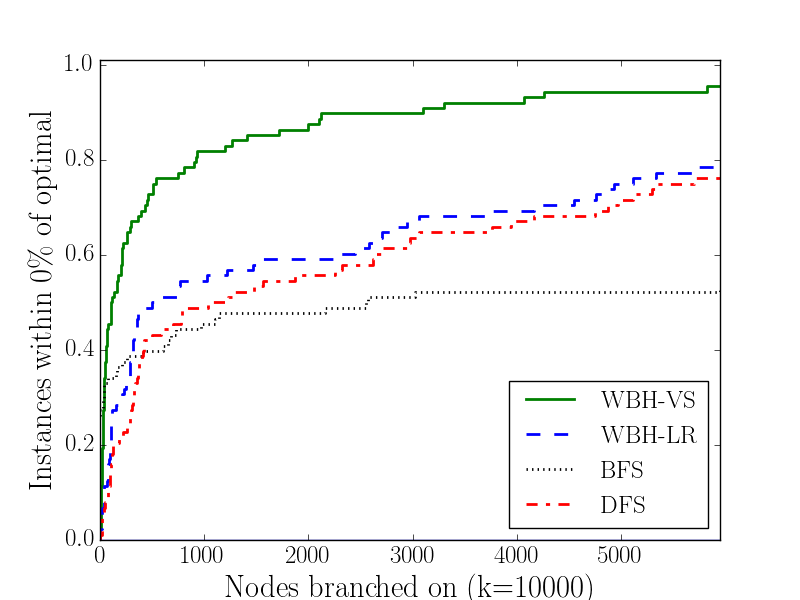}
\includegraphics[scale=.42]{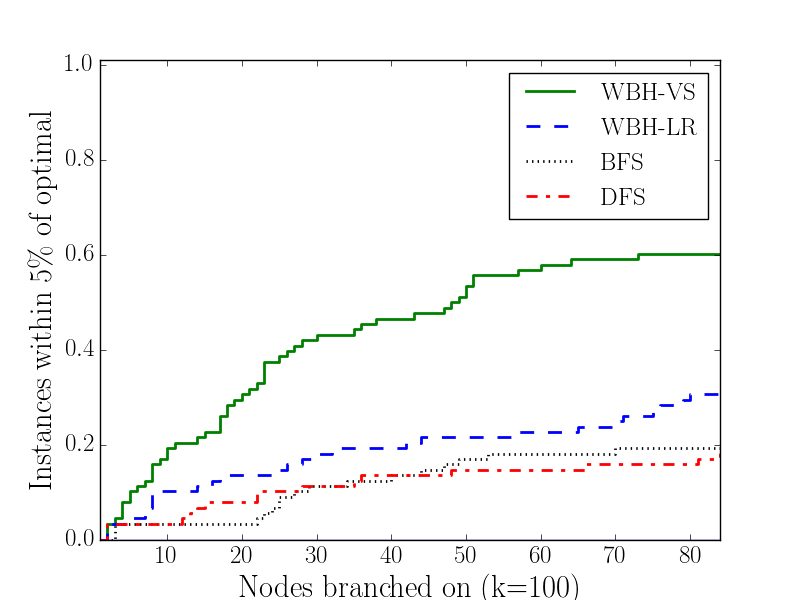}
\includegraphics[scale=.42]{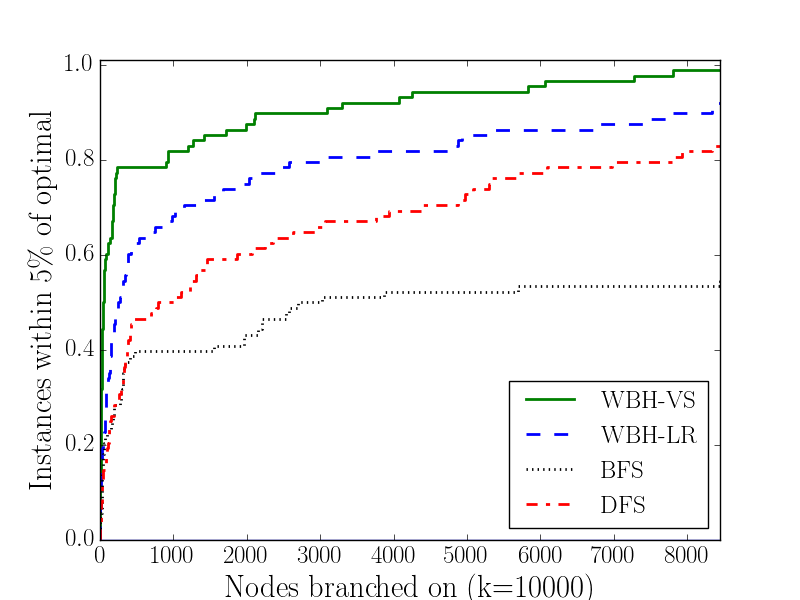}

\end{figure}
\begin{figure}[tb]
\caption{\revin{A performance profile showing the fraction of \texttt{Turner} instances with 30 vertices for which a target bound is proved.   The partial tree is limited to at most $100$, or $1\,000$ internal nodes. 
} }\label{fig:turner30}
\includegraphics[scale=.42]{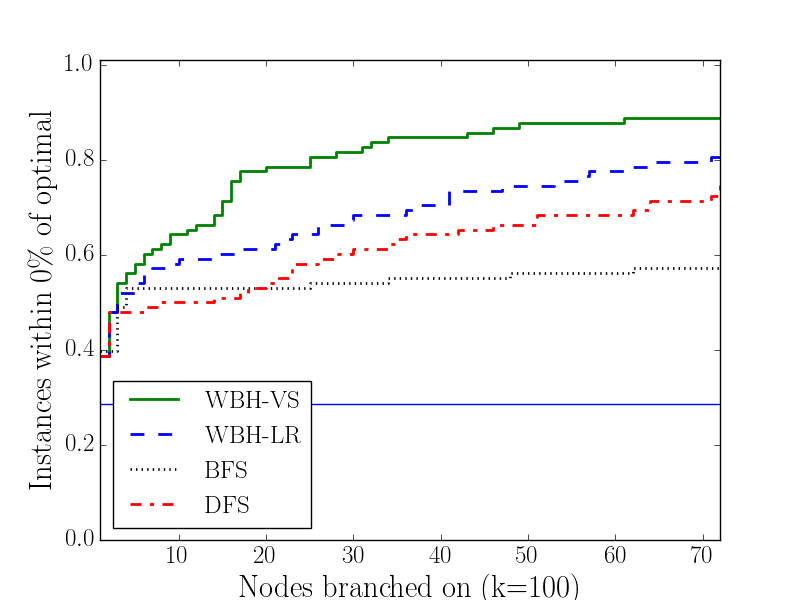}
\includegraphics[scale=.42]{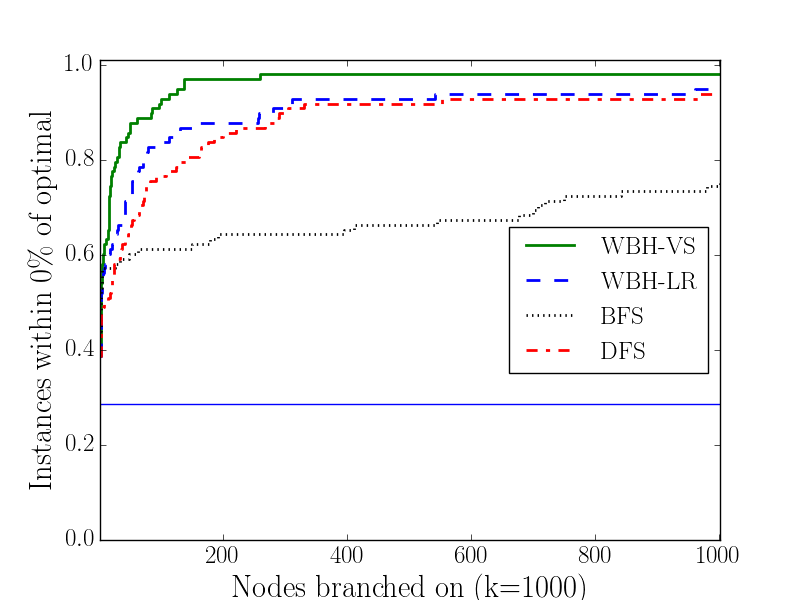}
\end{figure}
\begin{figure}[tb]
\centering
\caption{\revin{A performance profile showing the fraction of \texttt{Turner} instances with 100 vertices for which a target bound is proved. The partial tree is limited to at most   $1\,000$  internal nodes. }}\label{fig:turner100}
\includegraphics[scale=.4]{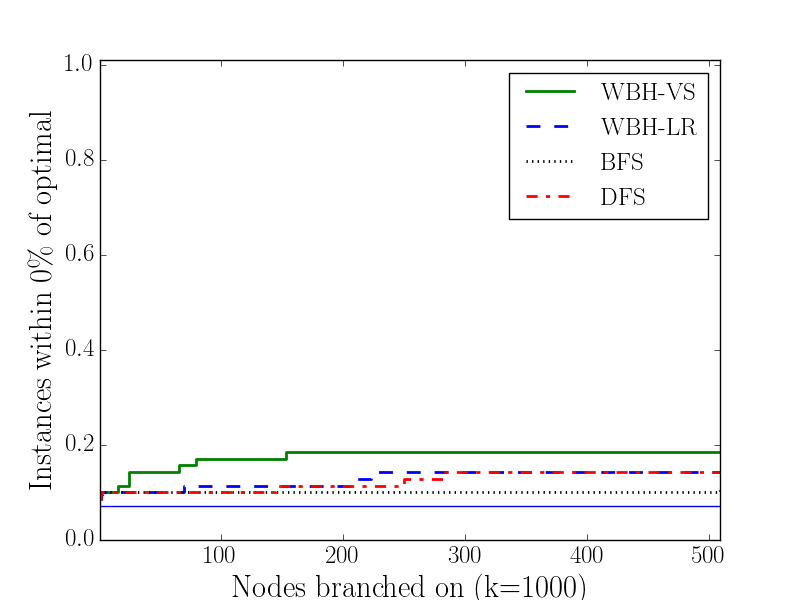}
\includegraphics[scale=.4]{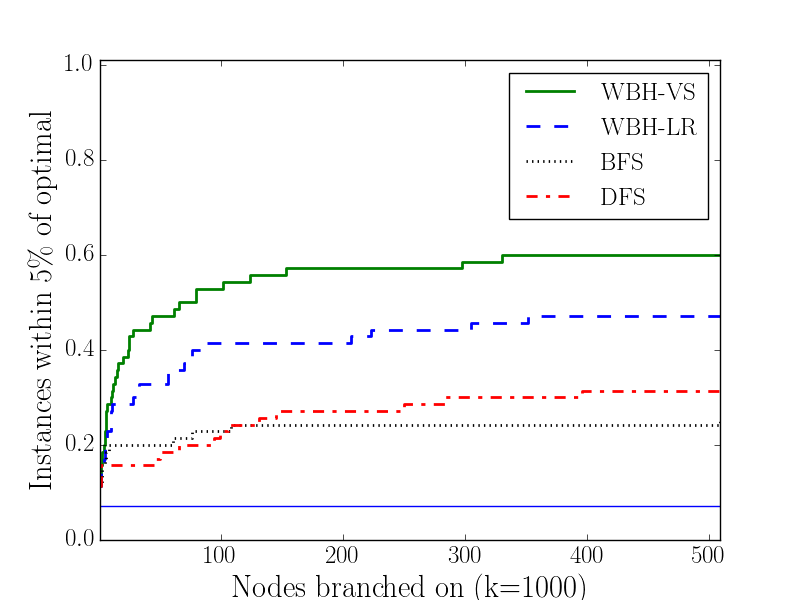}
\end{figure}

\revin{Figures~\ref{fig:rand30}, \ref{fig:turner30}, \ref{fig:turner100},\ref{fig:turner250} and \ref{fig:benchmarks} are a representative sample of performance profiles that indicate the fraction of instances (vertical axis) for which a given branching strategy proves a target bound after branching on a given number of nodes (horizontal axis). The target bound is typically  within 0\% or 5\% of the optimal bandwidth (for the \texttt{Random} and \texttt{MM} instances), or $\phi$ (for the \texttt{Turner} instances). The branching strategies compared are WBH with greedy variable selection (WBH-VS) and WBH with layered branching order which alternates between placing vertices in the left and right of the permutation \mbox{(WBH-LR)}, as well as DFS and BFS with the same alternating branching order. Where relevant the figures also show the  fraction of instances for which the tighter of the graph theoretic lower bounds   (\ref{eq:blumbound}) and (\ref{eq:csbound}) achieved the target bound.  The number of internal nodes are limited to $k \in \{100,1\,000,10\,000\}$; when the horizontal axis terminates before $k$  all of the curves are flat for a larger number of branches up to $k$. We omit results for the \texttt{Turner} instances with 1000 vertices, since the gap between the tightest graph theoretic lower bound and the upper bound $phi$ was found to be $0.35\%$ on average, leaving little room for improvement.
}

\revout{Figure~\ref{fig:rand30} shows the results for the \texttt{Random} instances.  Each plot is a performance profile that indicates the fraction of instances (vertical axis) for which a given branching strategy proves a target bound after a given number of branches (horizontal axis).  The target bound is within 0\%, 5\%, or 10\% of the optimum, where the optimum is obtained by solving a constraint programming model of the bandwidth problem.  The branching strategies are WBH with greedy variable selection (WBH-VS) and WBH with fixed branching order \mbox{(WBH-FB)}, as well as DFS and BFS with the same fixed branching order.  The graph-theoretic bounds failed to achieve any of the targets and therefore do not appear in the plots.  Two plots are presented for each target, in which the the number of internal nodes is limited to $k=1000$ or $k=10,000$, respectively. We omit results for the \emph{Turner1000} instances since it was that the average gap between $\phi$ and  the strongest graph-theoretic bound was on average only $0.35\%$, leaving little room for improvement. }

\begin{figure}[tb]
\caption{\revin{Scatter plot showing the relative performance of BFS and WBH-LR compared to WBH-VS on the 90  \texttt{Random} instances with 30 vertices after branching on 100 nodes.} }\label{fig:rand30:scatter}
\includegraphics[scale=.42]{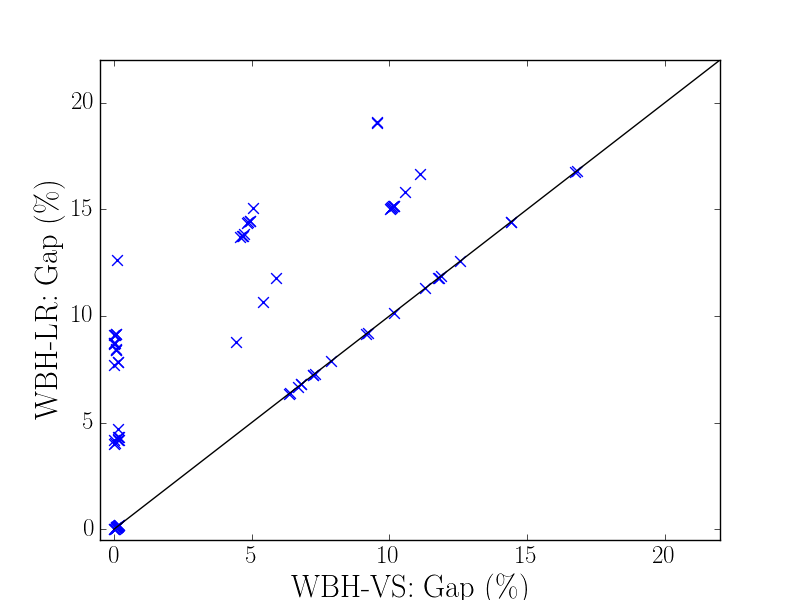}
\includegraphics[scale=.42]{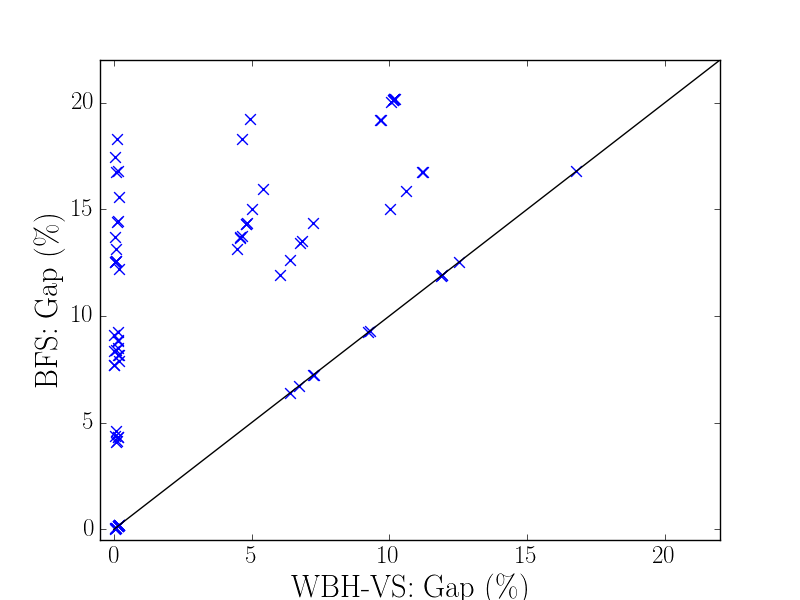}

\end{figure}

\begin{figure}[tb]
\centering
\caption{\revin{Scatter plot comparing WBH-VS with BFS and WBH-LR on the 70 \texttt{Turner} instances with 100 vertices after branching on $10\,000$ nodes.   }}\label{fig:turner100:scat}
\includegraphics[scale=.4]{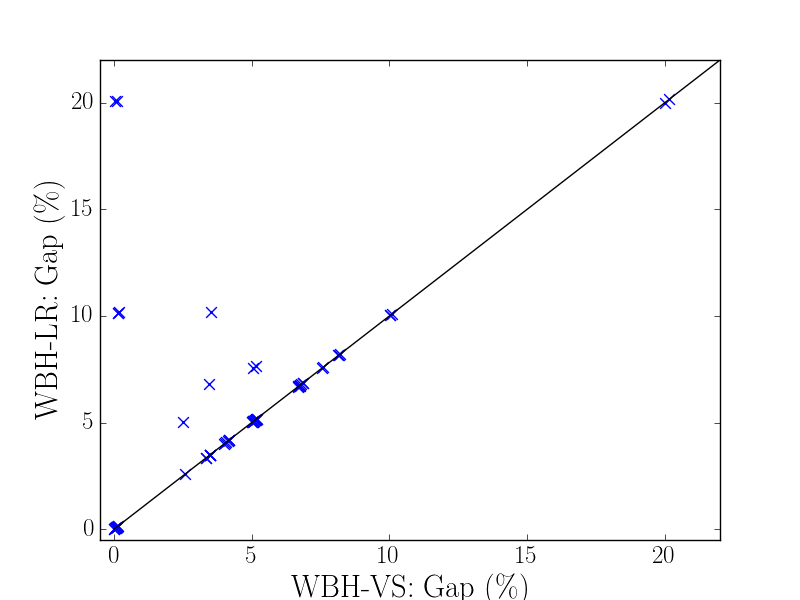}
\includegraphics[scale=.4]{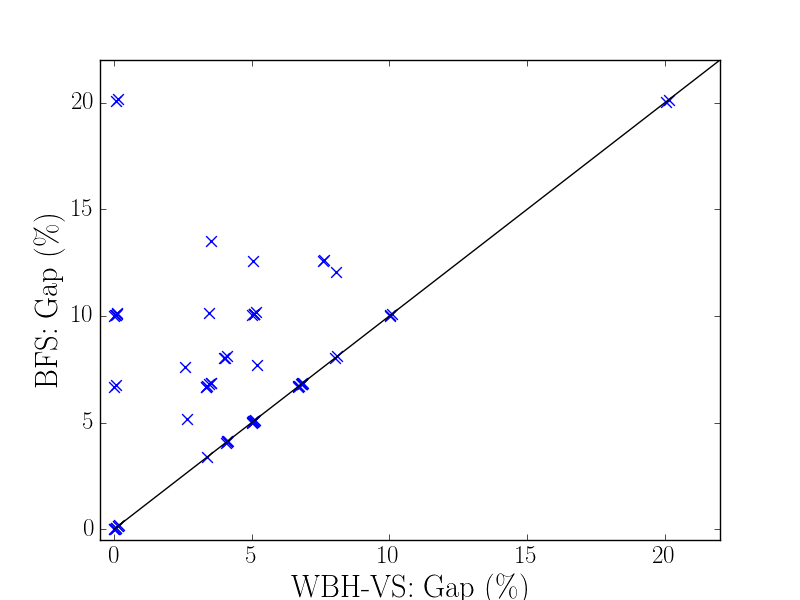}
\end{figure}

\begin{figure}[tb]
\centering
\caption{\revin{Scatter plot comparing WBH-VS with BFS and WBH-LR on the 39 \texttt{MM} instances after branching on $1\,000$ nodes.   }}\label{fig:benchmarks:scat}
\includegraphics[scale=.4]{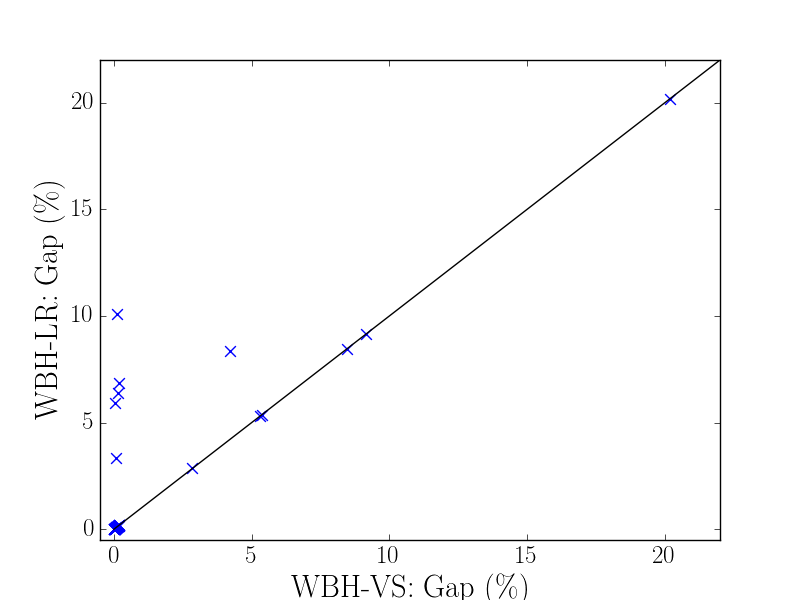}
\includegraphics[scale=.4]{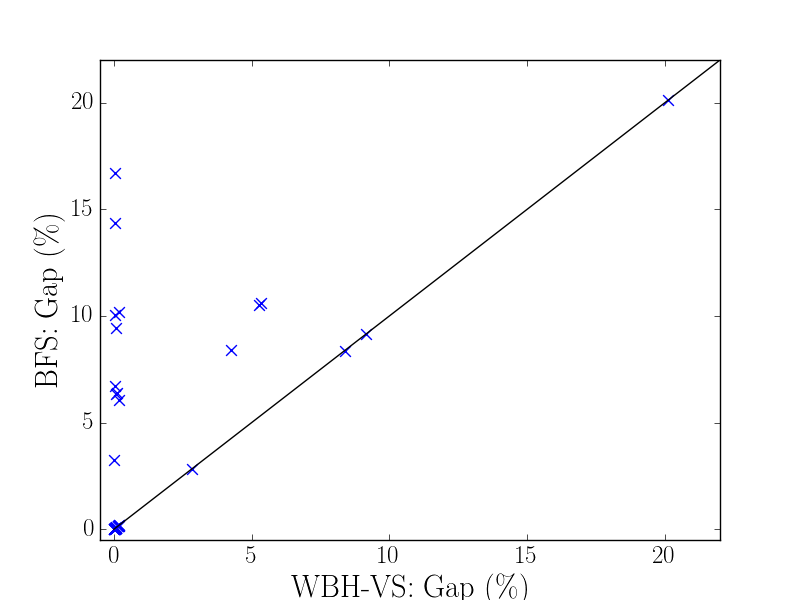}
\end{figure}

\revin{The figures indicate that worst-bound branching with fixed variable order is superior to both breadth-first and depth-first branching, and far superior to the graph-theoretic bounds.   Moreover, worst-bound branching with variable selection tightens the bound substantially, for any given time investment.  In fact, on the \texttt{Random}, \texttt{Turner30} and \texttt{MM} instances it obtains the optimal value for most instances after only modest computational effort. WBH-VS and WBH-LR retains a clear advantage over BFS and DFS on the larger \texttt{MM} instances. On the larger instances it is unsurprisingly more difficult for any of the branching methods to achieve the target bound in a limited number of nodes. 
On the   \texttt{Turner250} instances the worst-bound heuristic provides a smaller benefit over the alternatives, in part because the node limits are more restrictive in a large graph with an increased branching factor, and in part due of the fact that the gap between the graph theoretic lower bounds and $\phi$ is much smaller (see Table~\ref{tab:gap}). Yet the worst-first heuristic continues to prove stronger bounds than BFS and DFS while requiring significantly less computation.

Figures~\ref{fig:rand30:scatter}, \ref{fig:turner100:scat} and \ref{fig:benchmarks:scat} are scatter plots comparing the gap between the lower and upper bound for the respective branching strategies on a per-instance basis after branching on a specified number of nodes. The plots for the omitted test sets are similar.  We observe that WBH-VS inproves over WBH-LR and BFS on a significant fraction of the instances, including many which are solved optimally by WBH-VS while the alternative methods report gaps in excess of 5 or 10\%. 

Table~\ref{tab:gap} shows the average gap between the lower and upper bounds after branching on $100, 1\,000$ or $10\,000$ nodes for each of the methods. As benchmark we also report the average gap between the strongest of the graph theoretic bounds and the best known upper bound. For all test sets, the gap is  significantly reduced by WBH-LR and WBH-VS after branching on only 100 nodes. On all the test sets except \texttt{Turner100}, WBH-VS achieves an average gap of less than 1\% after branching on $10\,000$ vertices. We also observe that the graph-theoretic bounds are far tighter on the \texttt{Turner250} instances than any other, with an average gap of $4.7\%$. This decreases to $0.35\%$ in our randomly generated \texttt{Turner1000} instances (not shown). Finally, we remark that it is perhaps somewhat surprising that DFS on occasion reports smaller gaps than BFS. This suggests that having strong upper bounds is useful even when solving the branching dual. 

Table~\ref{tab:frontier} reports the maximum frontier size encountered while branching on up to $k\in \{100, 1\,000, 10\,000\}$ nodes, averaged over the instances in each of the test sets. The maximum frontier size is the largest number of open nodes stored at any point during the computation. As expected, DFS leads to significantly smaller frontier sizes than any of the other methods since it quickly probes to a layer deep in the tree where the average branching factor is likely to be much lower than at the root node, and spends the bulk of the computation time at that depth. The memory requirements of WBH-VS is  comparable to that of WBH-LR on the smaller instances, and within a factor of 2.5 on \texttt{Turner250} and \texttt{MM}. Both variants of the worst-bound heuristic generally have much smaller frontier sizes than BFS. 

Although our theoretical results do not guarantee the strongest bound for a given frontier size, we observe that the maximum frontier size is strongly correlated with the node limit $k$ and the size of the instance.  A user who has limited memory available may select a node limit $k$ appropriately and strengthen the bound as much as possible subject to the node limit $k$ as guaranteed by Corollary~\ref{cor:nodelimit}. This is unlikely to differ much from the best possible bound subject to an explicit constraint on the frontier size. 
}

\revout{
Figures~\ref{fig:turner30} and~\ref{fig:turner100} display results for the \textt{Turner} instances with 30 and 100 vertices, respectively.  Here the targets are within 0\%, 5\% and 10\% of $\phi$ rather the optimal value.  The horizontal baseline indicates the fraction of instances in which the tighter of the two graph theoretic bounds (\ref{eq:blumbound}) and (\ref{eq:csbound}) achieved the target.   }
\revout{
The results for 30 vertices are very similar to the \texttt{Random} results, except that the relative advantage of variable selection is slightly smaller.  In the results for 100 vertices, it is (unsurprisingly) more difficult for any of the branching methods to achieve the 0\% target.  Yet the worst-first heuristic continues to excel relative to BFS, DFS, and the graph-theoretic bounds.  }

\revout{
Figure~\ref{fig:benchmarks} represents the results on the \emph{Matrix Market} benchmark instances. For the instances where the optimal bandwidth is not known we compute the target bound in terms of the best known lower bound reported in \citet{CapSal05}. 
}

\revout{
On the larger \emph{Matrix Market} and \texttt{Turner} instances it is (unsurprisingly) more difficult for any of the branching methods to achieve the target bound in a limited number of nodes.  Yet the worst-first heuristic continues to excel relative to BFS, DFS, and the graph-theoretic bounds, and there remains some benefit it using the greedy lookahead variable selection strategy.
}

\revin{
\begin{figure}[tb]
\centering
\caption{\revin{A performance profile showing the fraction of \texttt{Turner} instances with 250 vertices for  which a target bound is proved. The partial tree is limited to at most $1\,000$, or $10\,000$  internal nodes. }}\label{fig:turner250}
\includegraphics[scale=.4]{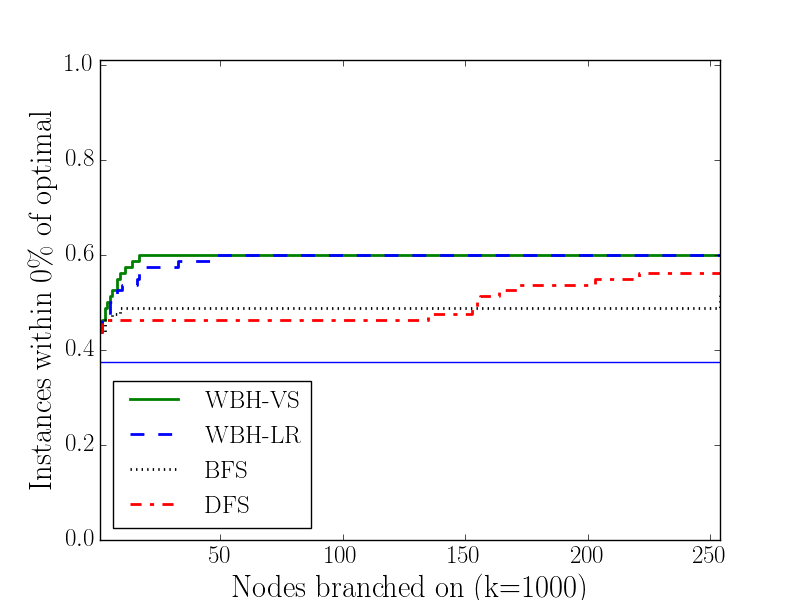}
\includegraphics[scale=.4]{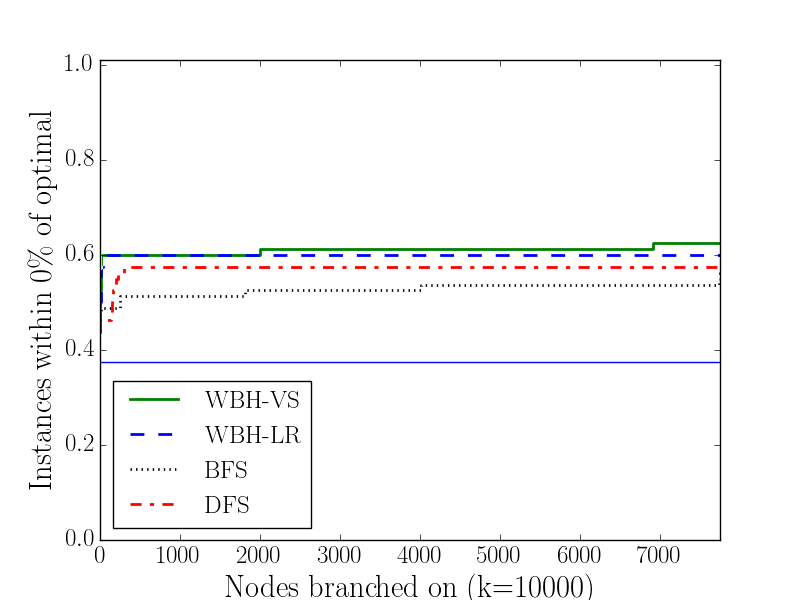}
\end{figure}
}

\revin{
\begin{figure}[tb]
\centering
\caption{\revin{A performance profile showing the fraction of \texttt{MM} instances  which a target bound is proved.  The partial tree is limited to at most $1\,000$, or $10\,000$  internal nodes. } }\label{fig:benchmarks}
\includegraphics[scale=.4]{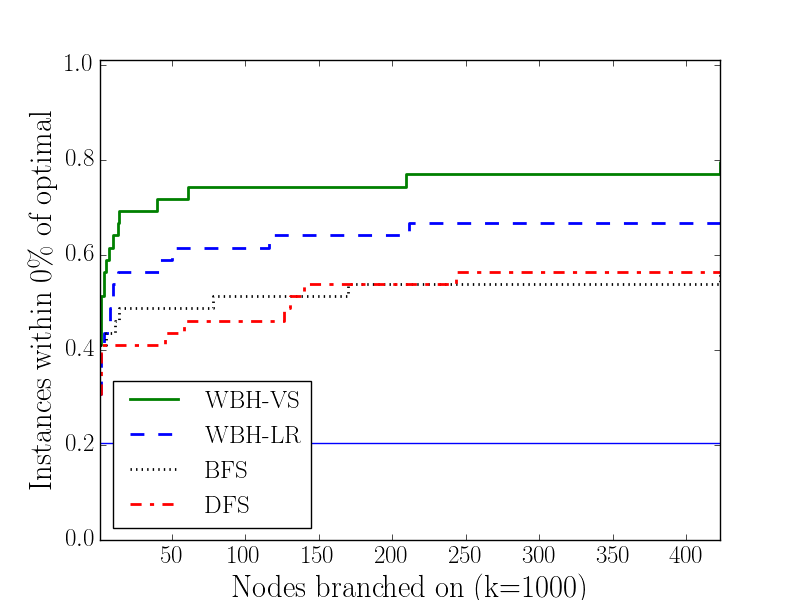}
\includegraphics[scale=.4]{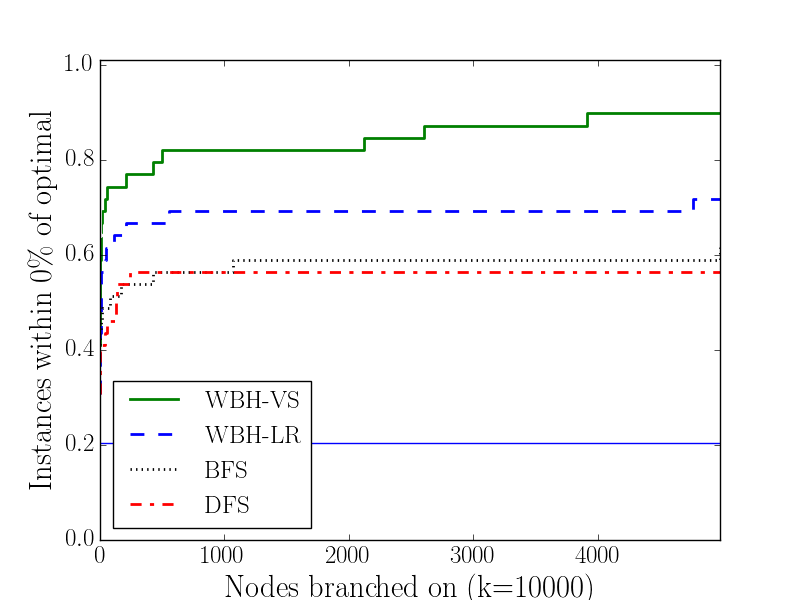}
\end{figure}
}

\begin{sidewaystable}[]
\centering
    \caption{\revin{Average gap size (\%) between the optimal lower bound and the best lower bound proved   after branching on $100$, $1\,000$ and $10\,000$ nodes on the respective test instances. Benchmark (BM) is the  strongest of the graph theoretic bounds. }}\label{tab:gap}
\begin{tabular}{lccccccccccccccc}
\toprule
  & \multicolumn{3}{c}{\texttt{Turner30}} & \multicolumn{3}{c}{\texttt{Turner100}} & \multicolumn{3}{c}{\texttt{Turner250}} &  \multicolumn{3}{c}{\texttt{Random30}} & \multicolumn{3}{c}{\texttt{Benchmarks}}  \\
\cmidrule(lr){2-4} \cmidrule(lr){5-7} \cmidrule(lr){8-10} \cmidrule(lr){11-13} \cmidrule(lr){14-16}
       & \multicolumn{3}{c}{Nodes branched on} & \multicolumn{3}{c}{Nodes branched on} & \multicolumn{3}{c}{Nodes branched on} & \multicolumn{3}{c}{Nodes branched on} & \multicolumn{3}{c}{Nodes branched on} \\
Method & 100 & $1\,000$ & $10\,000$ & 100 & $1\,000$ & $10\,000$ & 100 & $1\,000$ & $10\,000$ & 100 & $1\,000$ & $10\,000$ & 100 & $1\,000$ & $10\,000$  \\ \midrule
BFS     & 3.671 & 1.798 & 0.781 & 8.843 & 8.307 & 7.507 & 1.760 & 1.646 & 1.432  & 11.111 & 6.816 & 4.459& 5.099 & 4.054 & 3.503   \\
DFS     & 2.782 & 0.819 & 0.000 & 8.940 & 8.190 & 7.810  & 1.870 & 1.469 & 1.323 & 14.290 & 8.786 & 1.523& 5.937 & 4.630 & 4.630  \\
WBF-LR  & 1.497 & 0.564 & 0.000 & 6.871 & 6.100 & 5.707  & 1.214 & 1.068 & 1.016 & 8.098 & 3.350 & 0.723 & 3.368 & 2.334 & 2.067  \\
WBF-VS  & 0.654 & 0.213 & 0.000 & 5.933 & 5.505 & 4.457 & 1.068 & 0.964 & 0.880  & 4.307 & 1.429 & 0.076 & 1.926 & 1.406 & 0.283   \\
BM      & 27.99 & 27.99 & 27.99 & 18.07 & 18.07 & 18.07 & 4.729 & 4.729 & 4.729  & 32.53 & 32.53 & 32.53& 20.54 & 20.54 & 20.54  \\
\bottomrule
\end{tabular}\bigskip

    \caption{\revin{Average maximum frontier size after branching on $100$, $1\,000$ and $10\,000$ nodes on the test instances instances.  }}\label{tab:frontier}
\begin{tabular}{lccccccccccccccc}
\toprule
  & \multicolumn{3}{c}{\texttt{Turner30}} & \multicolumn{3}{c}{\texttt{Turner100}} & \multicolumn{3}{c}{\texttt{Turner250}} &  \multicolumn{3}{c}{\texttt{Random30}} & \multicolumn{3}{c}{\texttt{Benchmarks}}  \\
\cmidrule(lr){2-4} \cmidrule(lr){5-7} \cmidrule(lr){8-10} \cmidrule(lr){11-13} \cmidrule(lr){14-16}
       & \multicolumn{3}{c}{Nodes branched on} & \multicolumn{3}{c}{Nodes branched on} & \multicolumn{3}{c}{Nodes branched on} & \multicolumn{3}{c}{Nodes branched on} & \multicolumn{3}{c}{Nodes branched on} \\
       Method & 100 & $1\,000$ & $10\,000$ & 100 & $1\,000$ & $10\,000$ & 100 & $1\,000$ & $10\,000$ & 100 & $1\,000$ & $10\,000$ & 100 & $1\,000$ & $10\,000$  \\ \midrule
BFS    	&	557	&	 $2\,541$ 	&	 $7\,861$ 	&	 $2\,371$ 	&	 $22\,392$ 	&	 $184\,578$ 	&	 $4\,445$  	&	  $37\,049$ 	&	  $310\,281$ 	&	975	&	 $5\,800$ 	&	 $24\,699$ 	&	 $3\,794$   	&	 $28\,530$ 	&	 $211\,566$ \\
DFS    	&	 $82$ 	&	 $82$ 	&	 $82$     	&	 $753$   	&	 $753$   	&	 $753$      	&	 $2\,667$  	&	   $2\,880$  	&	   $2\,880$  	&	 $93$  	&	 $93$  	&	 $93$  	&	  $1\,640$	&	  $1\,794$	&	  $1\,794$\\
WBH-LR 	&	227	&	984	&	 $2\,968$ 	&	 $1\,058$ 	&	 $8\,461$ 	&	 $73\,035$ 	&	 $1\,306$  	&	   $10\,009$  	&	   $83\,734$  	&	391	&	 $2\,186$ 	&	 $8\,395$ 	&	 $1\,168$   	&	 $9\,667$  	&	 $78\,666$   \\
WBH-VS 	&	275	&	858	&	 $2\,041$ 	&	 $1\,926$ 	&	 $15\,318$ 	&	 $122\,814$ 	&	 $2\,570$  	&	 $18\,238$  	&	 $140\,324$  	&	573	&	 $2\,107$ 	&	 $6\,788$ 	&	 $3\,321$   	&	 $19\,643$	&	 $196\,432$\\

\bottomrule
\end{tabular}
\end{sidewaystable}

\section{Conclusion}

We studied the branching dual of an optimization problem for the purpose of strengthening an existing bound. We showed that for fixed variable selection, a natural worst-bound node selection heuristic is optimal for proving bounds with a given computational investment.  We evaluated the worst-bound heuristic experimentally on the minimum bandwidth problem using a relaxation function in \citet{CapLetSal11} and found that it is much more effective at proving bounds that depth-first or breadth-first search. Finally, we showed how combining this node selection heuristic with local search strategies for variable selection can lead to significant improvements in the quality of the bounds. 

The branching dual method is proposed here primarily for combinatorial problems that have no useful integer programming model.  In such cases, one need only determine how to strengthen a known bound, even a weak one, to reflect the fact that some variables have been fixed.  However, the same technique can also be used to obtain  bounds for integer \revin{or mixed integer} programming models\revout{ that are difficult to solve to optimality}.  In this case, modifying the linear programming bound to reflect fixed variables is trivial.  The method can also applied at individual nodes of a conventional branch-and-cut tree to obtain bounds that may be tighter than those obtained from a linear relaxation with cutting planes.  \revin{More generally, the method can be used for problems with a mixture of discrete variables (not necessarily integer) and continuous variables, so long as a relaxation value can be computed when some of the discrete variables have been fixed.}  These remain topics for future research.


%
%
%



\begin{thebibliography}{22}
	\providecommand{\natexlab}[1]{#1}
	\providecommand{\url}[1]{\texttt{#1}}
	\providecommand{\urlprefix}{URL }
	
	\bibitem[{Achterberg(2007)}]{Ach07}
	Achterberg T (2007) \emph{Constraint Integer Programming}. Ph.D. thesis,
	Technische {Universit\"{a}t} Berlin.
	
	\bibitem[{Achterberg et~al.(2005)Achterberg, Koch, \protect\BIBand{}
		Martin}]{AchKocMar05}
	Achterberg T, Koch T, Martin A (2005) Branching rules revisited.
	\emph{Operations Research Letters} 33(1):42--54.
	
	\bibitem[{Alvarez et~al.(2017)Alvarez, Louveaux, \protect\BIBand{}
		Wehenkel}]{AlvLouWeh17}
	Alvarez AM, Louveaux Q, Wehenkel L (2017) A machine learning-based
	approximation of strong branching. \emph{INFORMS Journal on Computing}
	29:185--195.
	
	\bibitem[{Applegate et~al.(2007)Applegate, Bixby, Chv\'{a}tal,
		\protect\BIBand{} Cook}]{AppBixChvCoo07}
	Applegate DL, Bixby RE, Chv\'{a}tal V, Cook WJ (2007) \emph{The Traveling
		Salesman Problem: A Computational Study} (Princeton University Press).
	
	\bibitem[{Benichou et~al.(1971)Benichou, Gautier, Girodet, Hentges, Ribiere,
		\protect\BIBand{} Vincent}]{BenGauGirHenRibVib71}
	Benichou M, Gautier JM, Girodet P, Hentges G, Ribiere R, Vincent O (1971)
	Experiments in mixed-integer linear programming. \emph{Mathematical
		Programming} 1:76--94.
	
	\bibitem[{Bixby et~al.(1995)Bixby, Cook, Cox, \protect\BIBand{}
		Lee}]{BixCooCoxLee95}
	Bixby RE, Cook W, Cox A, Lee EK (1995) Parallel mixed integer programming.
	Technical report {CRPC-TR95554}, Center for Research on Parallel Computation.
	
	\bibitem[{Blum et~al.(1998)Blum, Konjevodand, Ravi, \protect\BIBand{}
		Vempala}]{BluKonRavVem98}
	Blum A, Konjevodand G, Ravi R, Vempala S (1998) Semi-definite relaxations for
	minimum bandwidth and other vertex-ordering problems. \emph{Proceedings of
		the thirtieth annual ACM symposium on Theory of computing}, 100--105 (ACM).
	
	\bibitem[{Caprara et~al.(2011)Caprara, Letchford, \protect\BIBand{}
		Salazar-{Gonz\'{a}lez}}]{CapLetSal11}
	Caprara A, Letchford AN, Salazar-{Gonz\'{a}lez} JJ (2011) Decorous lower bounds
	for minimum linear arrangement. \emph{INFORMS Journal on Computing}
	23:26--40.
	
	\bibitem[{Caprara \protect\BIBand{} Salazar-Gonz{\'a}lez(2005)}]{CapSal05}
	Caprara A, Salazar-Gonz{\'a}lez JJ (2005) Laying out sparse graphs with
	provably minimum bandwidth. \emph{INFORMS Journal on Computing} 17:356--373.
	
	\bibitem[{Chv{\'a}tal(1970)}]{Chv70}
	Chv{\'a}tal V (1970) A remark on a problem of {Harary}. \emph{Czechoslovak
		Mathematical Journal} 20(1):109--111.
	
	\bibitem[{Dawande \protect\BIBand{} Hooker(2000)}]{DawHoo00}
	Dawande M, Hooker JN (2000) Inference-based sensitivity analysis for mixed
	integer/linear programming. \emph{Operations Research} 48:623--634.
	
	\bibitem[{Gautier \protect\BIBand{} Ribier(1977)}]{GauRib77}
	Gautier JM, Ribier R (1977) Experiments in mixed-integer linear programming
	using pseudo-costs. \emph{Mathematical Programming} 12:26--47.
	
	\bibitem[{Harvey \protect\BIBand{} Ginsberg(1995)}]{HarGin95}
	Harvey WD, Ginsberg ML (1995) Limited discrepancy search. \emph{IJCAI
		Proceedings}, 607--615.
	
	\bibitem[{Hooker(1996)}]{Hoo96}
	Hooker JN (1996) Inference duality as a basis for sensitivity analysis. Freuder
	EC, ed., \emph{Principles and Practice of Constraint Programming (CP 1996)},
	volume 1118 of \emph{Lecture Notes in Computer Science}, 224--236 (Springer).
	
	\bibitem[{Hooker(2012)}]{Hooker12}
	Hooker JN (2012) \emph{Integrated Methods for Optimization, 2nd ed.}
	(Springer).
	
	\bibitem[{Khalil et~al.(2016)Khalil, Bodic, Song, Nemhauser, \protect\BIBand{}
		Dilkina}]{KhaLeBSonNemDil16}
	Khalil EB, Bodic PL, Song L, Nemhauser G, Dilkina B (2016) Learning to branch
	in mixed integer programming. \emph{AAAI Proceedings}, 724--731.
	
	\bibitem[{Korf(1985)}]{Kor85}
	Korf RE (1985) Depth-first iterative-deepening: {An} optimal admissible tree
	search. \emph{Artificial intelligence} 27:97--109.
	
	\bibitem[{Linderoth \protect\BIBand{} Savelsbergh(1999)}]{LinSav99}
	Linderoth JT, Savelsbergh MWP (1999) A computational study of search strategies
	for mixed integer programming. \emph{INFORMS Journal on Computing}
	11:173--187.
	
	\bibitem[{Ostrowski et~al.(2011)Ostrowski, Linderoth, Rossi, \protect\BIBand{}
		Smriglio}]{OstLinRosSmr11}
	Ostrowski J, Linderoth J, Rossi F, Smriglio S (2011) Orbital branching.
	\emph{Mathematical Programming} 126:147--178.
	
	\bibitem[{Schulte et~al.(2017)Schulte, Tack, \protect\BIBand{}
		Lagerkvist}]{SchTacLag17}
	Schulte C, Tack G, Lagerkvist MZ (2017) Modeling and programming with gecode.
	User documentation.
	
	\bibitem[{Turner(1986)}]{Tur86}
	Turner JS (1986) On the probable performance of heuristics for bandwidth
	minimization. \emph{SIAM journal on computing} 15(2):561--580.
	
	\bibitem[{{Vil\'{i}m} et~al.(2015){Vil\'{i}m}, Laborie, \protect\BIBand{}
		Shaw}]{VilLabSha15}
	{Vil\'{i}m} P, Laborie P, Shaw P (2015) Failure-directed search for
	constraint-based scheduling. Michel L, ed., \emph{CPAIOR Proceedings}, volume
	9075 of \emph{Lecture Notes in Computer Science}, 437--453 (Springer).
	
\end{thebibliography}


\end{document}